\documentstyle[12pt,graphicx,pstricks,braket,amsmath,amssymb,enumitem,
                   float, epstopdf,pdfpages,cite,tabu,hyperref,subfigure, float,tikz]{article}
                    \usetikzlibrary{arrows,matrix,positioning}
\numberwithin{equation}{section}
\newcommand{\hi}[1]{}

\DeclareMathOperator{\Tr}{Tr}
\begin{document}

\def\AEF{A.E. Faraggi}

\def\JHEP#1#2#3{{JHEP} {\textbf #1}, (#2) #3}
\def\vol#1#2#3{{\bf {#1}} ({#2}) {#3}}
\def\NPB#1#2#3{{\it Nucl.\ Phys.}\/ {\bf B#1} (#2) #3}
\def\PLB#1#2#3{{\it Phys.\ Lett.}\/ {\bf B#1} (#2) #3}
\def\PRD#1#2#3{{\it Phys.\ Rev.}\/ {\bf D#1} (#2) #3}
\def\PRL#1#2#3{{\it Phys.\ Rev.\ Lett.}\/ {\bf #1} (#2) #3}
\def\PRT#1#2#3{{\it Phys.\ Rep.}\/ {\bf#1} (#2) #3}
\def\MODA#1#2#3{{\it Mod.\ Phys.\ Lett.}\/ {\bf A#1} (#2) #3}
\def\RMP#1#2#3{{\it Rev.\ Mod.\ Phys.}\/ {\bf #1} (#2) #3}
\def\IJMP#1#2#3{{\it Int.\ J.\ Mod.\ Phys.}\/ {\bf A#1} (#2) #3}
\def\nuvc#1#2#3{{\it Nuovo Cimento}\/ {\bf #1A} (#2) #3}
\def\RPP#1#2#3{{\it Rept.\ Prog.\ Phys.}\/ {\bf #1} (#2) #3}
\def\APJ#1#2#3{{\it Astrophys.\ J.}\/ {\bf #1} (#2) #3}
\def\APP#1#2#3{{\it Astropart.\ Phys.}\/ {\bf #1} (#2) #3}
\def\EJP#1#2#3{{\it Eur.\ Phys.\ Jour.}\/ {\bf C#1} (#2) #3}
\def\etal{{\it et al\/}}
\def\notE6{{$SO(10)\times U(1)_{\zeta}\not\subset E_6$}}
\def\E6{{$SO(10)\times U(1)_{\zeta}\subset E_6$}}
\def\highgg{{$SU(3)_C\times SU(2)_L \times SU(2)_R \times U(1)_C \times U(1)_{\zeta}$}}
\def\highSO10{{$SU(3)_C\times SU(2)_L \times SU(2)_R \times U(1)_C$}}
\def\lowgg{{$SU(3)_C\times SU(2)_L \times U(1)_Y \times U(1)_{Z^\prime}$}}
\def\SMgg{{$SU(3)_C\times SU(2)_L \times U(1)_Y$}}
\def\Uzprime{{$U(1)_{Z^\prime}$}}
\def\Uzeta{{$U(1)_{\zeta}$}}

\newcommand{\cc}[2]{c{#1\atopwithdelims[]#2}}
\newcommand{\bev}{\begin{verbatim}}
\newcommand{\beq}{\begin{equation}}
\newcommand{\ba}{\begin{eqnarray}}
\newcommand{\ea}{\end{eqnarray}}

\newcommand{\beqa}{\begin{eqnarray}}
\newcommand{\beqn}{\begin{eqnarray}}
\newcommand{\eeqn}{\end{eqnarray}}
\newcommand{\eeqa}{\end{eqnarray}}
\newcommand{\eeq}{\end{equation}}
\newcommand{\beqt}{\begin{equation*}}
\newcommand{\eeqt}{\end{equation*}}
\newcommand{\Eev}{\end{verbatim}}
\newcommand{\bec}{\begin{center}}
\newcommand{\eec}{\end{center}}
\newcommand{\bes}{\begin{split}}
\newcommand{\ees}{\end{split}}
\def\ie{{\it i.e.~}}
\def\eg{{\it e.g.~}}
\def\half{{\textstyle{1\over 2}}}
\def\nicefrac#1#2{\hbox{${#1\over #2}$}}
\def\third{{\textstyle {1\over3}}}
\def\quarter{{\textstyle {1\over4}}}
\def\m{{\tt -}}
\def\mass{M_{l^+ l^-}}
\def\p{{\tt +}}

\def\slash#1{#1\hskip-6pt/\hskip6pt}
\def\slk{\slash{k}}
\def\GeV{\,{\rm GeV}}
\def\TeV{\,{\rm TeV}}
\def\y{\,{\rm y}}

\def\l{\langle}
\def\r{\rangle}
\def\LRS{LRS  }

\begin{titlepage}
\samepage{
\setcounter{page}{1}
\rightline{LTH--1094}
\vspace{1.5cm}

\begin{center}
 {\Large \bf Axions And Self-Interacting Dark Matter \\In The Heterotic String-Derived Model}
\end{center}

\begin{center}
%\vspace{1.cm}

%\vfill 
{\large
Johar M. Ashfaque$^\spadesuit$\footnote{email address: jauhar@liv.ac.uk}
}\\
\vspace{1cm}
$^\spadesuit${\it  Dept.\ of Mathematical Sciences,
             University of Liverpool,
         Liverpool L69 7ZL, UK\\}
\end{center}

\begin{abstract}
After revisiting the heterotic string-derived low-energy effective model of \cite{Ashfaque:2016ydg, Athanasopoulos:2014bba, Faraggi:2016xnm, Ashfaque:2016jha} constructed in the four-dimensional free fermionic formulation, we find two axions which are either harmful or massive. As a direct consequence, they can not solve the strong $CP$ problem which is in complete agreement with \cite{Lopez:1990iq, Halyo:1993xn}. We also explore the possibility of the self-interacting dark matter residing in the non-Abelian gauge group present in the hidden sector \cite{Faraggi:2000pv}. We find that the low-energy string-derived model naturally welcomes the self-interacting dark matter as $4$ copies of the non-Abelian, hidden $SU(2)$ gauge group factor are present.
\end{abstract}
\smallskip}
\end{titlepage}

\section{Introduction}
Quantum chromodynamics (QCD) is a wonderful theory of the strong interactions. Having said that, however, it suffers from one serious problem: the strong $CP$ problem \cite{Cheng:1987gp}. There are three known viable solutions to tackle the strong $CP$ problem of which axions are the most plausible solution as they help to keep the strong $CP$ problem in check. 

The strong $CP$ problem emerges as a consequence of adding the $CP$ violating term to the QCD Lagrangian 
$${\mathcal{L}_{CP}}= \frac{\overline{\theta} \alpha_{s}}{32\pi^{2}}\tilde{G}_{\mu\nu}G^{\mu\nu}$$
which is a renormalizable and gauge invariant term that violates $CP$ and is allowed in any gauge theory in four dimensions. In the Standard Model (SM) it contributes to the $CP$-odd observables such as the nEDM.  The very smallness of  $\overline{\theta}$ despite large amounts of $CP$ violation in the weak sector of the SM is called the strong $CP$ problem. A feasible solution to the strong $CP$ problem,  \cite{Peccei:1977hh, Peccei:1977ur}, was proposed
by Peccei and Quinn (PQ) which later was dubbed as the ``axion" \cite{Weinberg:1977ma, Wilczek:1977pj}.

On the other hand, dark matter appears to make up about five times more mass than ordinary matter. Astrophysical and observational constraints indicate that dark matter is non-baryonic, cold and collisionless in nature. A picture that has come to be known as Cold Dark Matter (CDM). One of the major challenges is to understand the very nature of dark matter where ample evidence is provided by astrophysical observations. Dark matter could be made of a completely new, as yet undiscovered particle.

In this note, we present a study of axions in the string-derived low-energy effective model of \cite{Ashfaque:2016ydg, Athanasopoulos:2014bba, Faraggi:2016xnm, Ashfaque:2016jha}. We observe that there are two axions: the model-independent (MI) axion which is present in all superstring models \cite{Witten:1984dg} and is known to be massive \cite{Choi:1985je} while the PQ-type axion associated with the global anomalous $U(1)$ is present in all the models in the free fermionic formulation \cite{Lopez:1990iq, Halyo:1993xn}. We also explore the possibility of the self-interacting dark matter residing in the non-Abelian gauge group present in the hidden sector, in light of the heterotic string-derived low-energy effective model of \cite{Ashfaque:2016ydg, Athanasopoulos:2014bba, Faraggi:2016xnm, Ashfaque:2016jha}.

\section{A String-Derived Low-Energy Effective Model} 
The string-derived model in \cite{frzprime} was 
constructed in the free fermionic formulation \cite{fff} of the four-dimensional heterotic string. 
The details along with the the massless spectrum  and the superpotential can be found in \cite{frzprime} and are therefore omitted here. The string model as observed in \cite{frzprime} contains two anomalous $U(1)$s with $$\Tr U(1)_{1}=36,\qquad\qquad\Tr U(1)_{3}=-36$$ 
where the only anomalous linear combination was found to be $$U(1)_{A} = U(1)_{1}-U(1)_{3}$$ with $$\Tr{U(1)_A} = 72.$$  This anomalous combination, in turn, generates a Fayet-Iliopoulos $D$-term that breaks supersymmetry near the Planck scale \cite{dsw}. This model also contains vector--like states, displayed in table \ref{tablehi}, that transform under the hidden $SU(2)^4\times SO(8)$ group factors and therefore this model offers the possibility of accommodating the self-interacting dark matter candidates as was suggested in \cite{Faraggi:2000pv}.

\begin{table}[H]
	\begin{center}
		\begin{tabular}{|c|c|c|c|}
			\hline
			Symbol& Fields in \cite{frzprime} & ${SU(2)}^4\times SO(8)$&${U(1)}_{\zeta}$\\
			\hline
			$H^+$&$H_{12}^3$&$\left({\bf2},{\bf2},{\bf1},{\bf1},{\bf1}\right)$&$+1$\\
			&$H_{34}^2$&$\left({\bf1},{\bf1},{\bf2},{\bf2},{\bf1}\right)$&$+1$\\
			$H^-$&$H_{12}^2$&$\left({\bf2},{\bf2},{\bf1},{\bf1},{\bf1}\right)$&$-1$\\
			&$H_{34}^3$&$\left({\bf1},{\bf1},{\bf2},{\bf2},{\bf1}\right)$&$-1$\\
			$H$&$H_{12}^1$&$\left({\bf2},{\bf2},{\bf1},{\bf1},{\bf1}\right)$&$\,\,\,\,0$\\
			&$H_{13}^i,i=1,2,3$&$\left({\bf2},{\bf1},{\bf2},{\bf1},{\bf1}\right)$&$\,\,\,\,0$\\
			&$H_{14}^i,i=1,2,3$&$\left({\bf2},{\bf1},{\bf1},{\bf2},{\bf1}\right)$&$\,\,\,\,0$\\
			&$H_{23}^1$&$\left({\bf1},{\bf2},{\bf2},{\bf1},{\bf1}\right)$&$\,\,\,\,0$\\
			&$H_{24}^1$&$\left({\bf1},{\bf2},{\bf1},{\bf2},{\bf1}\right)$&$\,\,\,\,0$\\
			&$H_{34}^i,i=1,4,5$&$\left({\bf1},{\bf1},{\bf2},{\bf2},{\bf1}\right)$&$\,\,\,\,0$\\
			$Z$&$Z_i,i=1,\dots,$&$\left({\bf1},{\bf1},{\bf8}\right)$&$\,\,\,\,0$\\
			\hline
		\end{tabular}
	\end{center}
	\caption{\label{tablehi}
		Hidden sector field notation and associated states in \cite{frzprime}. }
\end{table}

\section{{The Spontaneous Breaking of  $U(1)_{A}$}}
The Dine-Seiberg-Witten anomaly cancellation mechanism at the field theory level \cite{dsw} is driven by the dilaton field which generates a large Fayet-Iliopoulos $D$-term for the $U(1)_A$ which would break supersymmetry thereby destabilizing the vacuum. However, one can assign VEVs along the flat directions to some $SO(10)\times U(1)_{\zeta}$ singlet fields in the massless spectrum of \cite{frzprime} to cancel the FI $D$-term and stabilize the vacuum by restoring supersymmetry.
Basically, we are looking to satisfy the inequality
\begin{center}
\framebox[2\width]
{$\sum_{i}Q_{A}^{i}|\langle\phi_{i}\rangle|^{2}<0.$} 
\end{center}
In general, it should be noted that all the local and global $U(1)$s will be spontaneously broken by the DSW mechanism.

The general form of the anomalous $D$-term associated to the $U(1)_{A}$ is 
$$D_{A} = \sum_{i}Q_{A}^{i}|\phi_{i}|^{2}+\frac{\alpha g^{2} e^{\Phi_{D}}}{192\pi^{2}}\Rightarrow \langle D_{A}\rangle=0= \sum_{i}Q_{A}^{i}|\langle \phi_{i}\rangle|^{2} +\frac{\alpha g^{2}}{192\pi^{2}}
$$ 
where $$\alpha\equiv \Tr U(1)_{A} = 72$$
which leads to the conclusion
$$\Tr U(1)_A  =  72 \Rightarrow \sum_{i}Q_{A}^{i}|\langle\phi_{i}\rangle|^{2} = - \frac{3 g^{2}}{8\pi^{2}}
 <0 .$$
The $U(1)_{A}$ is therefore spontaneously broken and to be more precise, it is broken by the VEVs of the large number of scalars \cite{Halyo:1993xn}. Consequently, one finds $$f_{a}\sim \frac{M}{10}\sim 10^{17}\,\,\mathtt{GeV}$$ where $M=M_{Pl}/2\sqrt{8\pi}$ which is deemed unacceptable since it has been known for a long time that a large axion decay constant, $f_{a}$, especially $f_{a}>10^{12}\,\,\mathtt{GeV}$, means axion energy density $\rho_{a}>\rho_{\text{critical}}$ \cite{Cheng:1987gp}.

\section{{Discussion on Self-Interacting Dark Matter}}
The following relation between $N_{f}$, $N_{C}$ and $h$ was obtained in \cite{Faraggi:2000pv} by setting $$M_{h}=10^{h}\,\,\text{GeV} $$ and taking $\Lambda_{h}=1\,\,$GeV$ $, $\alpha_{h}(\Lambda_{h}) = 1$ and $\alpha_{0}=\textstyle{\frac{1}{24}}$ 
$$\frac{1}{2}N_{f}(h-16)+48N_{C} = \frac{46\pi}{\ln 10}$$ 
beginning with one-loop RGE for the hidden gauge coupling 
$$\frac{1}{\alpha_{h}(\mu)} = \frac{1}{\alpha_{0}} -\frac{1}{2\pi}\bigg[\bigg(\frac{1}{2}N_{f}-3N_{C}\bigg)\ln \frac{M_{h}}{M_{S}}-3N_{C}\ln \frac{\Lambda_{h}}{M_{h}}\bigg]$$
where 
$N_{C} = n$ is the number of colors, $N_{f}$ is the number of spin-$\textstyle{\frac{1}{2}}$ fundamental multiplets and $\alpha_{0}$ is the value of the gauge coupling at $M_{S} = 10^{16}\,\,$GeV$ $ with the scale at which the hidden $SU(n)$ gauge group becomes strongly interacting is defined to be 
$$\Lambda_{h} = M_{S}\exp \bigg(\frac{2\pi (1-\alpha_{0})}{(\frac{1}{2}N_{f}-3N_{C})\alpha_{0}}\bigg).$$  It is given that the number of needed flavors will grow at an immense rate with incrementing $N_{C}$. Subsequently the conclusion is reached that the wisest choice for the gauge group should have the smallest number of colors, in other words, the $SU(2)$ gauge group. It was also shown in \cite{Faraggi:2000pv} that the standard-like models of \cite{Faraggi:1997dc, Faraggi:1989ka, Faraggi:1991jr, Faraggi:1991be} are very interesting due to the frequent emergence of the $SU(2)$ hidden gauge group factors.

\section{Conclusions}
In this note, we reviewed the string-derived low-energy effective model in the free fermionic formulation and found that the two axions were either harmful or massive. Therefore, they can not solve the strong $CP$ problem which is in complete agreement with \cite{Lopez:1990iq, Halyo:1993xn}. We have also explored the possibility of the self-interacting dark matter residing in the non-Abelian gauge group present in the hidden sector and found that the low-energy effective model naturally welcomes the self-interacting dark matter as $4$ copies of the non-Abelian, hidden $SU(2)$ gauge group factor are present.

\section{Acknowledgements}
J. M. A. would like to thank the String Phenomenology 2016 organisers hosted in Ioannina and String-Math 2016 organisers hosted in Paris for their warm hospitality.

\end{document}